\documentclass[12pt]{article}
\usepackage{sectsty}
\usepackage{mathrsfs}
\usepackage[nosort]{cite}
\usepackage[pdftex]{graphicx}
\usepackage[font={scriptsize,singlespacing}]{caption}
\usepackage{epstopdf}
\usepackage{amsmath}
\usepackage{amssymb}
\usepackage{subfigure}
\usepackage{hyperref}
\usepackage{url}
\usepackage{xcolor}
\usepackage{color}

\definecolor{palatinateblue}{rgb}{0.15, 0.23, 0.89}
\hypersetup{ linktoc=all,
    colorlinks, linkcolor={palatinateblue},
    citecolor={palatinateblue}, urlcolor={palatinateblue}
}

\def\sideremark#1{\ifvmode\leavevmode\fi\vadjust{\vbox to0pt{\vss
 \hbox to 0pt{\hskip\hsize\hskip1em
 \vbox{\hsize2cm\tiny\raggedright\pretolerance10000
 \noindent #1\hfill}\hss}\vbox to8pt{\vfil}\vss}}}%
\begin{document}
\thispagestyle{empty}
\begin{center}

\null \vskip-1truecm \vskip2truecm

{\Large{\bf {The Information Paradox for Black Holes. \\ \vskip0.2truecm }}}

\vskip1truecm

{{S. W. Hawking,}}\\
\vskip0.1truecm
{DAMTP, \\ Centre for Mathematical Sciences,\\  University of Cambridge, \\ Wilberforce Road,\\ Cambridge,  CB3 0WA\\ UK.}\\

\end{center}
\vskip1truecm \centerline{{ABSTRACT}} \baselineskip=15pt

\medskip
\begin{quote}
{I propose that the information loss paradox can be resolved by considering the supertranslation of the horizon caused by the ingoing particles. Information can be recovered in principle, but it is lost for all practical purposes.}
\end{quote}
\let\thefootnote\relax\footnote{Talk given on 28 August 2015 at \lq\lq Hawking Radiation\rq\rq,  a conference held at KTH Royal Institute of Technology, Stockholm.}
\newpage

Forty years ago I wrote a paper, ``Breakdown  of Predictability in Gravitational Collapse'' \cite{hawking1976}, in which I claimed there would be loss of predictability of the final state if the black hole evaporated completely. This was because one could not measure the quantum state of what fell into the black hole. The loss of information would have meant the outgoing radiation is in a mixed state and the S-Matrix was non-unitary.

Since the publication of that paper, the AdS/CFT correspondence has shown there is no information loss. This is the information paradox: How does the information of the quantum state of the infalling particles re-emerge in the outgoing radiation?
This has been an outstanding problem in theoretical physics for the last forty years. Despite a large number of papers (see reference \cite{amps,ampss} for a list), no satisfactory resolution has been found. I now propose that the information is stored, not in the interior of the black hole (as one might expect), but on its boundary, the event horizon. This is a form of holography.

The concept of supertranslations was introduced in 1962 by Bondi, Metzner and Sachs (BMS) \cite{BM1962,S1962}, to describe the asymptotic isometries of an asymptotically flat spacetime in the presence of gravitational radiation. In other words the BMS group describes the symmetry on $\mathscr{I}^+$. For an asymptotically flat spacetime,  a supertranslation $\alpha$ shifts the retarded time $u$ to
\begin{equation}
u' = u + \alpha,
\end{equation}
where $\alpha$ is a function of the coordinates on the 2-sphere. 
The supertranslation moves each point of $\mathscr{I}^+$ a distance $\alpha$ to the future along the null geodesic generators of $\mathscr{I}^+$.  Note that the usual time and space translations form a four parameter sub-group of the infinite dimensional supertranslations but they are not an invariant sub-group of the BMS group.

Listening to a lecture by Strominger on the BMS group, \cite{SZ}, at the Mitchell Institute for Fundamental Physics and Astronomy workshop this April, I realized that stationary black hole horizons also have supertranslations.   In this case, the advanced time $v$ is shifted by $\alpha$, that is,
\begin{equation}
v' = v + \alpha.  
\end{equation}
The null geodesic generators of the horizon need not have a common past end point and there is no canonical cross section of the horizon. The tangent vector $l$ to the horizon is taken to be normalized  such that it agrees with the Killing  vectors, of time translation and rotation, on the horizon.

Classically, a black hole is independent  of its past history.   I shall assume this is also true in the quantum domain.  How then can a black hole emit the information about the particles that fell in?  The answer I propose, as explained  above, is that the information is stored in a supertranslation  associated with the shift of the horizon that the ingoing particles caused. 

The supertranslations form a hologram of the ingoing particles. The varying shifts along each generator  of the horizon leave an imprint on the outgoing particles in a chaotic but deterministic manner. There is no loss of information.
Note that  although the discussion in this  paper focuses on the asymptotically flat case, this proposal also works for black holes on arbitrary backgrounds,  e.g., in the presence of a nonzero cosmological constant.

Polchinski recently used a shock wave approximation to calculate the shift on a generator of the horizon caused by an ingoing wave packet \cite{P2015}. Even though the calculation may require some corrections,  this shows in principle that the ingoing particles determine a supertranslation of the black hole horizon. This in turn, will determine varying delays in 
the emission of wave packets. The information about the ingoing particles is returned, but in a highly scrambled, chaotic and useless form. This resolves the information paradox. For all practical purposes, however, the information is lost.

Unlike the suggestion of 't Hooft,  \cite{thooft}-\cite{thooft1}, that  relies on a cut-off of high energy  modes near the horizon, the resolution of the information loss paradox  I proposed here is based on symmetries, namely supertranslation invariance of the S-matrix between the ingoing and outgoing particles scattered off the horizon, which by construction is unitary.

A full treatment of the issues presented here will appear in a future publication with M. J. Perry and A. Strominger, \cite{HPS}.


\end{document}